\documentclass[lettersize,journal]{IEEEtran}

\def\subparagraph{} 
\usepackage{titlesec}
\titlespacing*{\section}{0pt}{*1}{*1}
\titlespacing{\subsection}{0pt}{*1}{*1}

\renewcommand{\thesubsubsection}{\arabic{subsubsection}}

\titleformat{\subsubsection}[runin]{\itshape}{\thesubsubsection)}{1em}{}
\titlespacing*{\subsubsection}{\parindent}{0pt}{*1}

\usepackage{cite}
\usepackage{amsmath,amssymb,amsfonts,nccmath,siunitx}
\usepackage{svg}
\usepackage{textcomp}
\usepackage{xcolor}
\usepackage{textgreek}
\usepackage{booktabs,multirow,rotating}
\usepackage{threeparttable}
\usepackage{makecell}
\usepackage{orcidlink}
\usepackage{fancyhdr}
\usepackage{etoolbox}
\usepackage{mathtools}
\usepackage{float}
\usepackage{dblfloatfix} 
\usepackage{subcaption} 
\usepackage{graphicx}
\usepackage{bm}

\hypersetup{hidelinks} 

\usetikzlibrary{calc,angles,quotes,arrows.meta}

\tikzset{pill/.style={minimum width=1.2cm,minimum height=6mm,rounded
corners=3mm,draw},
reactor/.style={circle,draw,minimum size=6mm,path picture={
\draw (-3mm,0) -- (3mm,0) (0,-3mm) -- (0,3mm);
\fill (0,0) -- (3mm,0) arc(0:-90:3mm) -- cycle;
\fill (0,0) -- (-3mm,0) arc(180:90:3mm) -- cycle;
}}}

\begin{document}

\setlength{\abovedisplayskip}{3pt} 
\setlength{\belowdisplayskip}{3pt}

\title{
Traction and Stability Control using Fuzzy-based Controller Integration for Electric Vehicles}

\author{Nimantha Dasanayake,~\IEEEmembership{Student Member,~IEEE,} \orcidlink{0000-0001-6221-1610} and Shehara Perera,~\IEEEmembership{Member,~IEEE}\orcidlink{0000-0001-6695-4307} 
\thanks{\vspace{5pt}}
\thanks{Nimantha Dasanayake is with the Department of Mechanical Engineering, Faculty of Engineering,
        University of Moratuwa, Moratuwa, Sri Lanka
        {\tt\small dasanayakenp.21@uom.lk}}%
\thanks{Shehara Perera is with the Dyson School of Design Engineering,
        Imperial College London, United Kingdom,
        {\tt\small u.perera@imperial.ac.uk}
        }%
}

\maketitle
\thispagestyle{empty}
\pagestyle{empty}

\begin{abstract} 

Adverse road conditions can cause vehicle yaw instability and loss of traction. To compensate for the instability under such conditions, corrective actions must be taken. In comparison to a mechanical differential, an electronic differential can independently control the two drive wheels and provide means of generating more effective corrective actions. As a solution for traction and stability issues in automobiles, this study has developed a controller for a vehicle electronic differential consisting of two program-controlled rear motors. The control algorithm adjusts to changing road conditions. Traction was controlled using a motor reaction torque observer-based slip ratio estimation, and yaw stability was achieved by tracking a reference yaw rate calculated using estimated tyre cornering stiffnesses. A recursive least squares algorithm was used to estimate cornering stiffness. The yaw rate of the vehicle, as well as its longitudinal and lateral accelerations, were measured, and the body slip angle was estimated using an observer. A fuzzy inference system was used to integrate the independently developed traction control and yaw control schemes. The fuzzy inference system modifies the commanded voltage generated by the driver's input to account for the traction and yaw stability controller outputs. A vehicle simulator was used to numerically simulate the integrated controller. For a racetrack simulation, the peak slip ratio was reduced by 42.31\% and the RMS yaw rate error was reduced by 88.17\%. For the double lane change test at 40 \unit{km.h^{-1}} and 100 \unit{km.h^{-1}}, the RMS yaw rate error was reduced by 86.96\% and 92.34\%, respectively.

\end{abstract}

\maketitle

\section{Introduction}
    \label{Introduction}
\vspace{-3pt}
\IEEEPARstart{T}{he} study of controlling vehicle dynamics to produce dependable and safer rides has recently evolved dramatically. More emphasis has been placed on creating Electronic Differentials with Traction and Stability Control (EDTSCs). These technologies are utilised to manage the longitudinal and lateral dynamics of the vehicle, making Electric Vehicles (EVs) safer and more robust in operation. Because the driver's input and the control signals from the Traction and Stability Control (TSC) system must be carefully combined to have stable control over the drive train, the requirements for an EDTSC system for an EV are more demanding and sophisticated than those for internal combustion engine vehicles. This complexity, however, opens up new options for improving control over the vehicle's nonlinear dynamics.
 
Commercial EV manufacturers have used technologies such as torque vectoring, EDTSCs, and TSCs to improve traction and stability during critical driving situations and ensure optimal driving performance \cite{[2],[3],[4]}.

\begin{figure}[tbp]
\begin{center}
 \centering
\includegraphics[]{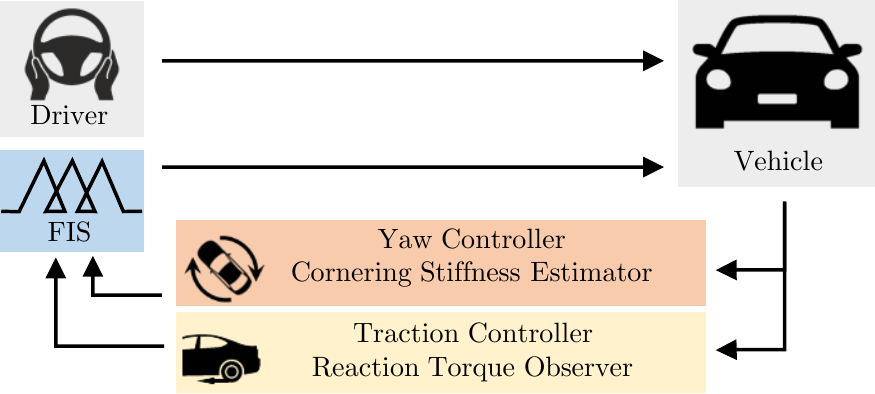}
\vspace{-15pt}
\caption{Simplified representation of the proposed control architecture for the EDTSC with Fuzzy Inference System (FIS) and yaw and traction controllers.}
\vspace{-25pt}
\label{Intro - Simpliified Architecture}
\end{center}
\end{figure}

\section{Related Work}
    \label{Related Work}

A commonly used method for vehicle yaw stability control in electronic differentials is direct yaw-moment management \cite{[7],[15],[25],[50]}. The system independently adjusts the driving and braking forces on each wheel to generate an inverse yaw moment to counteract the yaw moments that cause instability in crucial situations under direct yaw-moment control. To prevent wheel spin or skid caused by saturation of longitudinal tyre force, direct torque control of motors has been developed singly or together with direct yaw-moment management. 

\begin{table*}[tbp]
  \centering
  \caption{Traction and yaw moment control algorithms}
  \begin{threeparttable}
    \renewcommand{\arraystretch}{0.8}
    \setlength{\tabcolsep}{0.5em}
    \begin{tabular}{@{}p{10em}p{22em}p{20em}p{8.555em}}
    \toprule
    Control Objectives & State Measurements/ Estimations & Control Algorithm & Disadvantages \\
    \midrule
    Yaw rate & Yaw rate, (In addition: Slip angle observer, Cornering stiffness estimators) & LQR \cite{[7]}, FIS \cite{[35]}, Adaptive Neuro-FIS \cite{[41]}, SMC \cite{[9]}, Adaptive-SMC \cite{[21]}, Feedback and adaptive feedforward control \cite{[25],[50]}, MPC \cite{[16]} & No traction control incorporated\\
    \midrule
    Slip ratio & Wheel speed, (In addition: Slip ratio,  Friction coefficient estimators) & SMC \cite{[26],[27]}, Disturbance observer based current control \cite{[30]} & No yaw control incorporated\\
    \midrule
    Yaw rate and Slip ratio & Yaw rate, Slip angle observer, Cornering stiffness estimator, Wheel speed (In addition: Slip ratio estimator)  & Feedback control \cite{[15]} & No yaw and traction control integration\\
    \bottomrule
    \end{tabular}%
    LQR: Linear Quadratic Regulator, FIS: Fuzzy Inference System, SMC: Sliding Mode Control, MPC: Model Predictive Control
    \end{threeparttable}
  \label{tab:control_algorithms}%
  \vspace{-15pt}
\end{table*}%

\subsection{State Estimation}	

To assess the amount of stability and hence conduct corrective actions, it is critical to perceive the dynamical states of the vehicle such as vehicle speed, accelerations, yaw rate, and wheel speeds. Direct torque and speed calculations of electric motors are one of the advantages of EVs over IC engine-driven vehicles in terms of control. A prominent control tool utilised by many researchers for states that are not immediately quantifiable is state estimator. Some examples are body slip angle observer \cite{[10]}, cornering stiffness and road friction coefficient estimator \cite{[13]} and, yaw-moment observer \cite{[15]}. Along with the above estimators, techniques like the Kalman filter \cite{[25]} and disturbance observer \cite{[30]} have been used to reduce the noise and uncertainty of the estimated values.

Calculation of slip ratio using the vehicle velocity incorporates the long time integral errors since the only way to estimate the velocity is to integrate the measured longitudinal acceleration. Therefore, in this study instead of using vehicle speed to directly calculate slip ratio an indirect way of estimating slip ratio that is based on reaction torque observer was used. The reaction torque observer is a disturbance observer which helps to achieve a robust estimation of slip ratio. The disturbance observer is widely used for estimating plant uncertainties and external disturbances \cite{[67]}. A brief study of the disturbance observation will be offered herein to elucidate the theory behind the reaction torque observer.

\subsection{Disturbance Observer}
 
For robust control against system disturbances, disturbance observer-based controllers are used in applications such as servo systems and robot manipulation. A deliberately designed disturbance observer deduces robust control, which compensates for the disturbance by appropriate feedback. According to \cite{[64]}, a disturbance observer was used to measure the reaction torque of joint actuators on a robotic manipulator. To improve system dynamics and stability by reducing response time and torque ripples against load disturbances, a torque disturbance observer has been implemented for model-predictive control of electric drives \cite{[69]}.

Although most recent work has used a disturbance observer to compensate for the effects of uncertainties on closed-loop control, the proposed method has only used it to realise the reaction torque under the assumption that model coefficient variations from nominal values are negligible.

\subsection{Control Strategies}

Moving beyond the standard approaches to dealing with the problem of vehicle dynamics control, the methodologies listed below have been added to the control architectures.

\subsubsection{Model Matching Control (MMC)}
A simple algorithm for developing a desired model of a dynamic system and matching the real states of the system with the desired model. The most typical states used by this type of algorithm are slip ratio, body slip angle, and yaw rate \cite{[23]}. 

\subsubsection{Optimal Control}
Used to determine the optimal values for control signals. The most common approach is to use the optimal control method to determine the control yaw moment by minimising a performance index or cost function composed of errors in body slip angle and yaw rate relative to desired/reference values \cite{[15]}.

\subsubsection{Model Predictive Control (MPC)}
As an extension of model matching control, the optimal path to match the dynamical states to the reference states is predicted and the control signal is updated accordingly. In terms of definition, MPC is a strategy that employs the theory of optimal control \cite{[24],[25]}.

\subsubsection{Sliding Mode Control (SMC)}
Limits the dynamic system state trajectory to a sliding surface with the desired characteristics and ensures that the system converges to the surface in a finite time \cite{[26]}. This approach is widely used in vehicle traction and/or stability control algorithms to deal with dynamical uncertainty \cite{[27],[29],[30]}. One drawback of SMC is that the control state trajectory chatters when it approaches closer to the controlled parameter's set point. This difficulty is overcome by utilising a solution that uses a conditional integrator to transit from SMC to PI controller when the parameter is close to the set point \cite{[6]}. 

\subsubsection{Fuzzy Logic and Neural Network Control} Have been utilised to improve the robustness of control algorithms in traction control, stability control, or both. Neural networks have been utilised as sub-algorithms within the main algorithm, which is in charge of a localised control \cite{[39],[38]}. Fuzzy logic control has been adapted in car traction control in a variety of ways \cite{[12],[35],[9],[41],[21]}. Adaptive fuzzy systems has also been used to integrate two different vehicle models with different cornering stiffnesses to obtain a yaw moment controller which is more robust to variations of the road surface \cite{[41]}. 

Table \ref{tab:control_algorithms} offers a summary of the control algorithms used in previous studies, along with their disadvantages. The primary drawback observed by the authors is the absence of either traction control, yaw moment control, or their coalition in action. It is challenging to integrate an EDTSC's traction and yaw stability control counterparts together. The essential issue is how to ensure that longitudinal dynamics are not compromised by efforts to control lateral dynamics. In previous work, fuzzy systems hasn't been employed to address the issue of controller integration for a combined traction and yaw moment control. This work offers a controller integration method that uses a Fuzzy Inference System (FIS) to create corrective yaw moments utilising the slip ratio and error in yaw rate compared to the desired yaw rate as inputs. Furthermore, a motor reaction torque observer-based slip ratio estimation is proposed while employing the Recursive Least Square (RLS) algorithm for cornering stiffness estimation (see Fig. \ref{Intro - Simpliified Architecture}). 

More precisely and compactly, the contributions of this article are as follows:

\begin{enumerate}
    \item Estimation of the slip ratio indirectly using the disturbance observer applied for reaction torque observance.
  \item Using a fuzzy inference system for traction and yaw rate controller integration.
\end{enumerate}

The rest of the article is structured as follows. Section \ref{Vehicle Dynamics Model} presents the vehicle dynamics model used to develop the EDTSC. The Section \ref{EDTSC System Development} discusses the mathematical formulation of the proposed EDTSC. Section \ref{Experiments} shows the simulation model and experiments. Section \ref{Results} presents results and Section \ref{Conclusion} concludes the findings and presents the future work.

\section{Vehicle Dynamics Model}
    \label{Vehicle Dynamics Model}

The vehicle status variables must be determined before the EDTSC system can be implemented. The yaw rate can be measured directly with an Inertial Measurement Unit (IMU) sensor, but the side slip angle cannot be measured and must be approximated using a vehicle mathematical model. The yaw stability controller is based on the vehicle's lateral motion dynamics. Because of its simplicity, a two-degrees-of-freedom (2-DoF) linear vehicle model is used to calculate the vehicle lateral motion state variables.

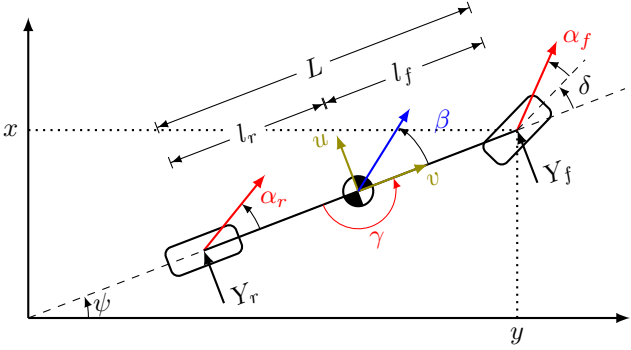
\begin{figure}[tbp]
\begin{center}
\begin{tikzpicture}

\pgfmathsetmacro{\y}{6.5}
\pgfmathsetmacro{\x}{2.5}
\pgfmathsetmacro{\Angle}{atan2(\x,\y)}
\pgfmathsetmacro{\AngleR}{30}
\pgfmathsetmacro{\AngleF}{45}
\pgfmathsetmacro{\AngleDelta}{25}
\pgfmathsetmacro{\AngleBeta}{37}
\pgfmathsetmacro{\COMradius}{0.2}

\coordinate (Origin) at (0,0);
\draw [-latex,thick] (Origin)--++(0,\x+1.5) coordinate (yaxis); 
\draw [-latex,thick] (Origin)--++(\y+1.5,0) coordinate (xaxis); 
\draw [dashed] (Origin)--++(\Angle:8.5cm) coordinate (AngleEnd);
\draw [dotted,thick] (\y,0) node [below] {$y$} --++(0,\x) coordinate (Yf);
\draw [dotted,thick] (0,\x) node [left] {$x$} --++(\y,0);
\draw pic["$\psi$", draw=black, text=black, -latex, angle eccentricity=1.25, angle radius=0.8cm]
              {angle=xaxis--Origin--Yf};

\coordinate (Yr) at ($ (Origin) + (\Angle:2.5cm) $);
\node at (Yr) [rotate=\Angle,draw,thick,rounded corners=1mm,minimum width=1cm, minimum height=0.4cm] {};
\draw [red,-latex,thick] (Yr)--++(\Angle+\AngleR:1.3cm) coordinate (RedArrowOne);
\draw pic["$\alpha_r$", draw=black, text=red, -latex, angle eccentricity=1.45, angle radius=0.8cm]
              {angle=Yf--Yr--RedArrowOne};
\draw [latex-,thick,black] (Yr)--++(\Angle-90:0.75cm) node [rotate=\Angle,right] {$Y_{r}$};           

\draw [thick] (Yr)--(Yf); 

\node at (Yf) [rotate=\Angle+\AngleDelta,draw,thick,rounded corners=1mm,minimum width=1cm, minimum height=0.4cm] {};         
\draw [red,-latex,thick] (Yf)--++(\Angle+\AngleF:1.3cm) coordinate (RedArrowTwo);
\draw [dashed] (Yf)--++(\Angle+\AngleDelta:1.3cm) coordinate (DeltaAngleEnd);
\draw pic["$\delta$", draw=black, text=black, -latex, angle eccentricity=1.35, angle radius=0.8cm]
           {angle=AngleEnd--Yf--DeltaAngleEnd};
\draw pic["$\alpha_f$", draw=black, text=red, -latex, angle eccentricity=1.45, angle radius=1cm]
      {angle=DeltaAngleEnd--Yf--RedArrowTwo};
\draw [latex-,thick,black] (Yf)--++(\Angle-90:0.75cm) node [rotate=\Angle,right] {$Y_{f}$};       

\coordinate (COM) at ($ (Origin) + (\Angle:4.7cm) $);
\begin{scope}[rotate=\Angle]
\fill [radius=\COMradius] (COM) -- ++(\COMradius,0) arc [start angle=0,end angle=90] -- ++(0,-2*\COMradius) arc [start angle=270, end angle=180];
\draw [thick,radius=\COMradius] (COM) circle;
\end{scope}
\draw [-latex,thick,olive] (COM)--++(\Angle+90:0.8cm) node [left,rotate=\Angle] {$u$};
\draw [-latex,thick,olive] (COM)--++(\Angle:1cm) node [below,rotate=\Angle] {$v$};
\draw pic["$\gamma$", draw=red, text=red, -latex, angle eccentricity=1.4, angle radius=0.5cm]{angle=Yr--COM--Yf};
\draw [blue,-latex,thick] (COM)--++(\Angle+\AngleBeta:1.3cm) coordinate (BlueArrowBeta);
\draw pic["$\beta$", draw=black, text=blue, -latex, angle eccentricity=1.45, angle radius=1.0cm]
      {angle=Yf--COM--BlueArrowBeta};

\coordinate (LrLabel) at ($ (Yr) +  (\Angle+90:1.25cm) $);
\coordinate (COMLabel) at ($ (COM) +  (\Angle+90:1.25cm) $);
\coordinate (LfLabel) at ($ (Yf) +  (\Angle+90:1.25cm) $);
\coordinate (LlrLabel) at ($ (Yr) +  (\Angle+90:1.75cm) $);
\coordinate (LlfLabel) at ($ (Yf) +  (\Angle+90:1.75cm) $);

\draw [{Bar}{latex}-{latex}{Bar}] (LrLabel)--(COMLabel) node [midway,sloped,fill=white] {$l_r$};
\draw [{Bar}{latex}-{latex}{Bar}] (COMLabel)--(LfLabel) node [midway,sloped,fill=white] {$l_f$};
\draw [{Bar}{latex}-{latex}{Bar}] (LlrLabel)--(LlfLabel) node [midway,sloped,fill=white] {$L$};

\end{tikzpicture}
\vspace{-5pt}
\caption{Geometric representation of the bicycle model}
\vspace{-20pt}
\label{Bicycle Model}
\end{center}
\end{figure}

As shown in Figure \ref{Bicycle Model}, the vehicle was modelled as the equivalent half-vehicle. In the figure, $\beta$ is the body slip angle and $\gamma$ is the yaw rate. $u$ and $v$ are the vehicle's lateral and longitudinal velocities respectively. $\delta$ is the steering angle and $\alpha_f$ and $\alpha_r$ are the front and rear tire side slip angles. $Y_f$ and $Y_r$ are the lateral tire forces on the front and rear wheels, respectively. $l_f$ and $l_r$  are the distances between the center of mass and front and rear axles respectively.

The dynamics of the lateral motion of a vehicle can be represented as,

\begin{equation}
    Mv \left (\frac{\mathrm{d\beta}}{\mathrm{dt}}+\gamma \right) = 2(Y_f+Y_r)
    \label{eq:21}
\end{equation}

\begin{equation}
    I \frac{\mathrm{d\gamma}}{\mathrm{dt}} = N_z-N_t
    \label{eq:22}
\end{equation}

where $I$ is the vehicle’s inertia around the vertical axis. ${N_z}$ is the control moment which is generated by the torque differences between the left and right in-wheel motors and ${N_t}$ is the moment which is generated by the lateral forces. Other terms are the same as the way described in the preceding sections. Furthermore, $N_t$ can be represented as, 

\begin{equation}
    N_t=2C_f \left (\beta+\frac{l_f}{v}\gamma-\delta \right)l_f - 2C_r \left (\beta-\frac{l_v}{v}\gamma \right)l_r
    \label{eq:23}
\end{equation}

where $C_f$ and $C_r$ are the cornering stiffnesses of the front and rear tyres, which are defined as the gradients of the lateral tyres force against the slip angle. For a typical tyre, cornering stiffness increases linearly with slip angle for a particular range before peaking, causing tyre forces to be saturated. Cornering stiffnesses are treated as constants in linearized dynamics, assuming that the vehicle is only operated inside the linear region of the slip angle.

The yaw moment controller's objective is to generate a corrective yaw moment on the vehicle using differential motor torques in oversteer and understeer situations. The corrective yaw moment is generated to maintain the desired yaw rate. The required yaw rate is determined by the vehicle's longitudinal velocity and steering angle. The yaw rate of the vehicle should satisfy the following equation during steady-state cornering:

\begin{equation}
    \gamma_{des} = \frac{v}{(L+Kv^2)}\delta
    \label{eq:24}
\end{equation}

where L is the wheelbase and K is the stability factor given by,

\begin{equation}
    K=\frac{M}{2L} \left (\frac{l_r}{C_f} -\frac{l_f}{C_r} \right )
    \label{eq:25}
\end{equation}

If $K$ is negative, the car will oversteer, while if $K$ is positive, the vehicle will understeer. As the equation \eqref{eq:25} shows, $C_f$ and $C_r$ are required to calculate the desired yaw rate required by the yaw moment controller. The section  \ref{EDTSC System Development} discusses its estimation.

\section{EDTSC Model Architecture}
    
\label{EDTSC System Development}

\begin{figure}[tbp]
\begin{center}
\centering
\includegraphics[width=0.5\textwidth]{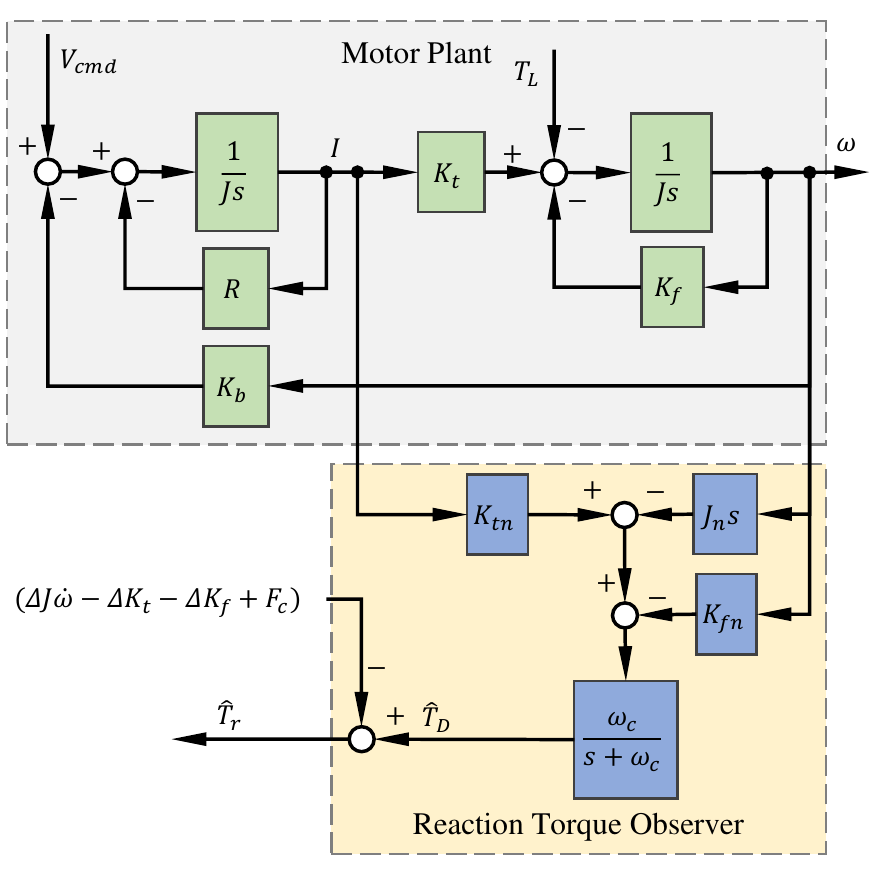}
\vspace{-15pt}
\caption{ Schematic diagram for motor plant with the reaction torque observer.}
\vspace{-25pt}
\label{ReactionT}
\end{center}
\end{figure}

The EDTSC proposed in this article is an adaptive model for controlling both traction and yaw stability. First, a disturbance observer-based reaction torque observer for an electric motor with current and rotor speed measurements is established. This facilitates an accurate estimation of the driving force acting on the wheels, resulting in more accurate direct torque control. The reaction torque observer formulation will be discussed next, followed by the direct torque control system, direct yaw moment control system, differential action, and FIS for integrating the traction and yaw stability control signals. 

\subsection{Reaction Torque Observer}

The equations of the motor model can be expressed as follows:

\begin{equation}
\mathrm{J\dot{\omega_m}} = {K_t}I-{K_f}{\omega_m}-{T_L}
\label{eq:1}
\end{equation}

\begin{equation}
\mathrm{L\dot{I}} = -RI-{K_b}{\omega_m} +{V_{cmd}} 
\label{eq:2}
\end{equation}

Here, 
$\mathrm{J} = {J_m}+{J_w}/G^{2}$ \label{eq:3}
is the combined inertia of the rotor ($J_m$) and the wheel ($J_w$) and $G$ is the gear ratio between the motor and the drive shaft. 
In equations \eqref{eq:1} and \eqref{eq:2}, $\omega_m$ is the motor speed, $I$ is the motor current, $T_L$ is the motor load torque, $V_{cmd}$ is the commanded voltage and,  $R$ and $L$ are the resistance and the inductance of the rotor winding.  $K_t$, $K_f$, $K_b$ are the torque, friction and back-EMF constants. The reaction torque ($T_R$) and the Coulomb friction ($F_c$) can be used to calculate motor load torque as
${T_L} = {F_c}+{T_R}$
\label{eq:4}.

The nominal motor model for velocity dynamics can be expressed as follows.

\begin{equation}
\mathrm{{J_n}\dot{\omega}_m} = {K_{tn}I+{K_{fn}}\omega_m -{T_D}}
\label{eq:5}
\end{equation}

The nominal combined inertia is $J_n$, the nominal torque and friction constants are $K_{tn}$ and $K_{fn}$, and the disturbance torque is $T_D$. It is assumed that current dynamics disturbances (equation \eqref{eq:2}) are insignificant. As a result, for the nominal plant transfer function ($P_n(s)$) that will be utilised to observe the disturbance, only the velocity dynamics are considered.

\begin{figure}[t]
\begin{center}
 \centering
\includegraphics[width=0.5\textwidth]{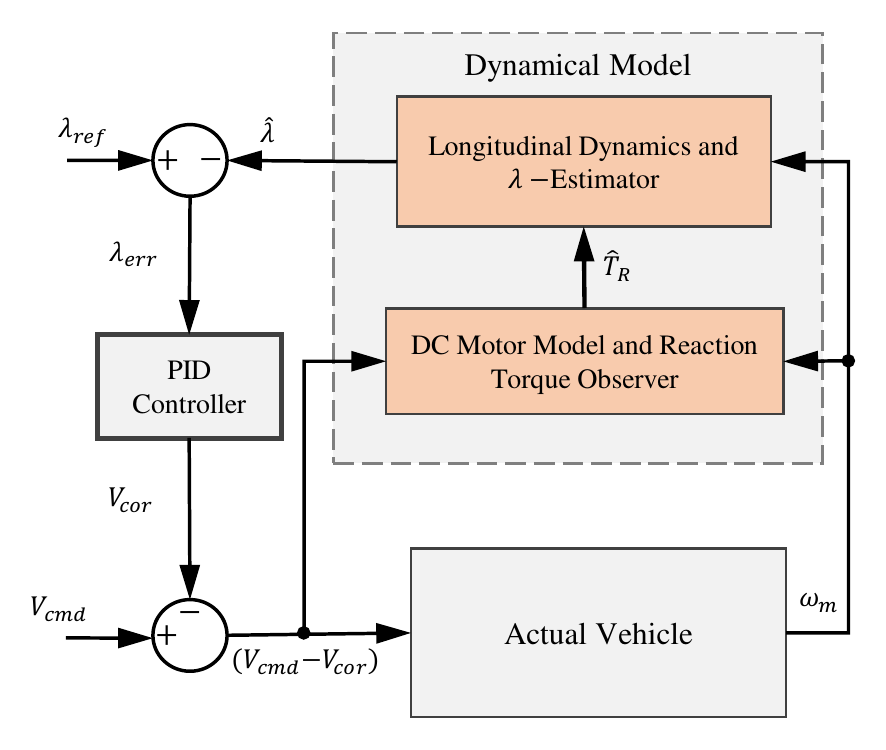}
\vspace{-20pt}
\caption{Schematic diagram of the direct torque controller.}
\vspace{-20pt}
\label{DTC}
\end{center}
\end{figure}

\begin{equation}
\mathrm{{P_n}(s)} = \frac{\omega_{m}(s)}{I(s) } = \frac{{K_{tn}}}{{J_n}s+{K_{fn}} }
\label{eq:6}
\end{equation}

The inertia of the motion system will change depending on its mechanical arrangement. Because of the irregular distribution of magnetic flux on the rotor's surface, the torque coefficient varies with rotor position in an electric motor. As a result, the disturbance torque is a combination of load torque and fluctuations in the constants $J_n$, $K_{tn}$, and $K_{fn}$. The constants with uncertainty can be written as $J = {J_n}+\Delta J$, ${K_t} = {K_{tn}}+\Delta{K_t}$ and ${K_f} = {K_{fn}}+\Delta{K_f}$. By substituting these constants to equation \eqref{eq:1}, and comparing with the equation \eqref{eq:5}, equation for disturbance torque can be derived as:

\begin{equation}
\mathrm{{T_D}} = \Delta J\dot{\omega_m}-\Delta {K_t}I - \Delta {K_f}\omega_m+{T_L}
\label{eq:10}
\end{equation}

From the equation \eqref{eq:5}, $T_D$ can be expressed as,

\begin{equation}
\mathrm{{T_D}} =  {K_{tn}}I - {K_{fn}}\omega_m -{J_n} {\dot{\omega}_m}
\label{eq:11}
\end{equation}

For several reasons, the equation  \eqref{eq:11} cannot be used directly to calculate $T_D$. To begin, when the calculation is implemented in a real system, the inverse of the nominal plant's transfer function (${P_{n}}^{-1} (s)$), which is not proper, must be implemented to obtain the result of the equation \eqref{eq:11}. However, improper transfer functions are not physically implementable \cite{[63]}. Second, the equation incorporates possibly noisy measurements $I$ and $\omega_m$. Third, the disturbance's direct feedback results in an algebraic loop. Furthermore, the time derivative of $\omega_m$ may increase the level of noise.

A first-order low-pass filter was used to overcome these concerns with a direct calculation of $T_D$  \cite{[63]}. The so-called Q-filter must be a low-pass filter with a relative degree greater than that of $P_n(s)$ and a wide bandwidth, thus a high cut-off frequency. A first-order filter with a cut-off frequency of $\omega_c$ is used in this work.

\begin{equation}
\mathrm{Q(s)} = \frac{{\omega_c}}{s + {\omega_c} }
\label{eq:12}
\end{equation}

The Q-filter is then used to get an estimate of  ${\hat{T}_D}$ as follows:

\begin{equation}
\mathrm{{\hat{T}_D}} = {\frac{{\omega_c}}{s + {\omega_c} }}.{({K_{tn}I-{K_{fn}\omega (s)-s{J_n}\omega (s))}}}
\label{eq:13}
\end{equation}

Overall motor dynamics with the disturbance torque observer is depicted in Figure \ref{ReactionT}. 

\begin{figure*}[!h]
\begin{center}
 \centering
 \vspace{-10pt}
\includegraphics[width=1\textwidth]{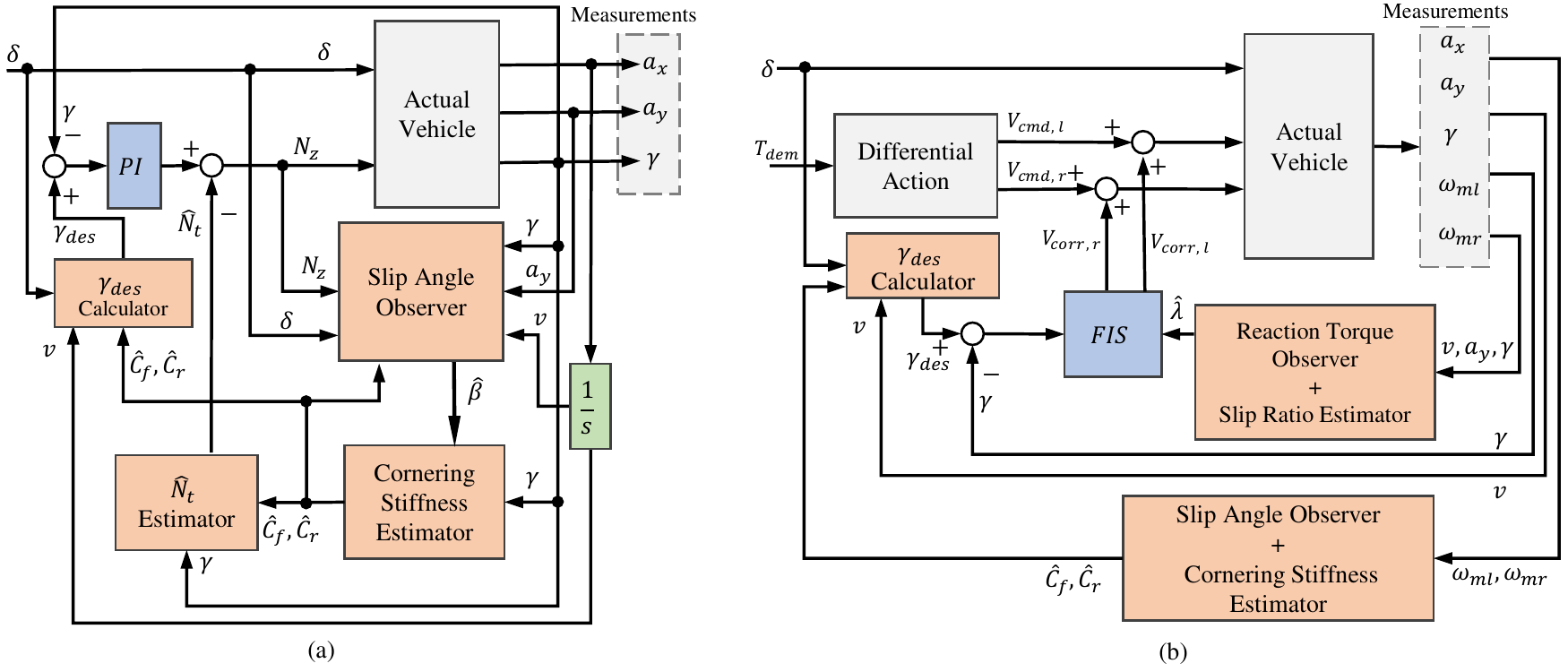}
\vspace{-10pt}
\caption{Controller schematics: (a) yaw stability controller, (b) traction and stability controller with fuzzy inference system (FIS) for controller integration.}
\vspace{-25pt}
\label{TSC}
\end{center}
\end{figure*}

The motor constants, namely, $J$, $K_t$, $K_f$, and $F_c$ can be determined using system identification and it was assumed that their variation during the operation is negligible. Therefore, from the equation \eqref{eq:10} and motor load torque expression, the estimated reaction torque can be derived as:

\begin{equation}
\mathrm{\hat{T}_D} \simeq {T_L} = {F_c}+{\hat{T}_R} 
\label{eq:14}
\end{equation}

\begin{equation}
\mathrm{\hat{T}_R} \simeq {\hat{T}_D}-{F_c} 
\label{eq:15}
\end{equation}

\subsection{Direct Torque Control}

The wheel slip ratio is a significant parameter used in traction control to avoid wheel skid and spin. To compute the slip ratio value of each wheel, the vehicle speed and wheel speed must be known. The vehicle speed is measured in conventional cars by the speeds of the non-driven wheels. Non-driven wheels slip in various conditions, such as braking and cornering over tight turns, and are thus unsuitable for calculating vehicle speed. An accelerometer cannot avoid the problem of speed offset caused by the long-time integral calculation \cite{[15]}. Based on the method provided in \cite{[15]}, a slip-ratio estimator was constructed to acquire an accurate wheel slip ratio value without detecting wheel speeds. However, unlike the original method, the slip ratio is calculated based on the reaction torque that was calculated by the reaction torque observer described in the previous section.

The longitudinal motion equation of the vehicle can be expressed as:

\begin{equation}
    M\dot{v} = F_d-F_{dr}  
    \label{eq:16}
\end{equation}

Where, $F_d$ is the driving force, $F_{dr}$ is the driving resistance, and $M$ is the vehicle mass. The relationship between the driving force and the reaction torque of the motor can be written using the wheel radius $r$ as $F_{dr}=T_R/r$. The wheel speed ($\omega$) can be obtained using the gear ratio of the gearbox ($G$) and motor speed $\omega_m$ as $G\omega_m = \omega$.

Slip ratio ($\lambda$) is a value that represents the difference between the vehicle speed and the tangential speed of the tire surface (${V_w}$) and is defined as:

\begin{equation}
    \lambda=(V_w-V)/V_w 
    \label{eq:19}
\end{equation}

where, ${V_w}=r\omega$. By differentiating the equation \eqref{eq:19} and substituting from above expressions, state equation for $\lambda$ is derived as: 

\begin{equation}
    \dot{\lambda} = \frac{\dot{\omega}_m}{\omega_m} \lambda + \frac{\dot{\omega}_m}{\omega_m} - \frac{\hat{T}_R}{r^2 M G \omega_m} + \frac{F_{dr}}{r M G \omega_m}
    \label{eq:20}
\end{equation}

The motor torque is regulated based on the reference slip ratio values to achieve the highest traction force at the tyre contact patch of each wheel. A PID controller can be used to manage motor torque by adjusting the commanded voltage by slip ratio error, as shown in Figure \ref{DTC}. However, a problem arises when the suggested direct torque controller is used in conjunction with the direct yaw moment controller. The control signals provided to the motor must respect both controllers; therefore, achieving a model-based optimal control strategy is challenging. To address this issue, FIS is proposed to generate corrective factors for motor voltage commands that account for yaw rate error as well as slip ratio, as stated in the following section.

\subsection{Direct Yaw Moment Control}
Since tyre cornering stiffnesses vary greatly depending on road conditions and vehicle load transfer, it is desirable to devise a method for estimating them. In previous research, a cornering stiffness estimator based on the Recursive Least Square (RLS) algorithm was utilised. The RLS algorithm presented in \cite{[25]} calculates the yaw moment created by lateral forces ($N_t$) using the measured yaw rate and estimated body slip angle, whereas the equations of front and rear lateral forces ($Y_f$, $Y_r$) are employed in \cite{[65]}. The equations used in \cite{[65]} were employed in this work, but the consideration of a parabolic relationship between slip and cornering stiffness was eliminated since the vehicle in this analysis is assumed to be operating in the linear region of slip angle.

Front and rear lateral forces ($Y_f$, $Y_r$) can be derived as:

\begin{equation}
    Y_f = M_y a_{yf} = M \frac{l_r}{L} (v(\hat{\dot\beta} + \gamma)+l_f \dot\gamma)\cos (\delta)
    \label{eq:26}
\end{equation}

\begin{equation}
    Y_r = M_y a_{yr} = M \frac{l_f}{L} (v(\hat{\dot\beta} + \gamma) - l_r \dot\gamma)
    \label{eq:27}
\end{equation}
	
Here, $\hat{\dot{\beta}}$ is the estimated body slip angle and its derivation will be discussed shortly. Now the regression equations can be constructed as follows:

\begin{equation}
    \bm{Y} = \bm{\xi}\bm{\theta}
    \label{eq:28}
\end{equation}

where,
\begin{equation}
    \mathbf{Y}^T = \begin{bmatrix} \hat{Y_f} & \hat{Y_r} \end{bmatrix}\hspace{10pt}
    \bm{\theta}^T = \begin{bmatrix} C_f & C_r \end{bmatrix}
    \label{eq:29}
\end{equation}

are the estimates of lateral forces and cornering stiffnesses respectively. The regressor $\bm{\xi}$ was derived as:

\begin{equation}
    \bm{\xi}(s) = F(s) \begin{bmatrix}
-2(\beta+\mfrac{l_f}{v}\gamma - \delta ) & 0\\
0 & -2(\beta - \mfrac{l_r}{v}\gamma)
\label{eq:30}
\end{bmatrix}
\end{equation}

where, $F(s)=\omega_c/(s+\omega_c)$ is a low pass filter used to filter out the noise of the time derivative terms by converting the transfer function into a proper one.  The estimation update is obtained from the following equation.

\begin{equation}
    \bm{\hat{\theta}}(k) = \bm{\hat{\theta}}(k-1)+\bm{K}(\bm{Y} - \bm{\xi}(k)\bm{\hat{\theta}}(k-1))
    \label{eq:31}
\end{equation}

The optimum RLS gain was calculated such that the squared error through time is minimized.

\begin{equation}
    \bm{K} = \bm{\Gamma}(k-1) \bm{\xi}^T(k) (\bm{R}+\bm{\xi}(k) \bm{\Gamma}(k-1) \bm{\xi}^T(k))^{-1}
    \label{eq:32}
\end{equation}

where, $\bm{\Gamma}$ is the estimation error covariance matrix and $\bm{R}$ is the measurement error covariance. An update of $\bm{\Gamma}$, is obtained as follows.

\begin{equation}
    \bm{\Gamma}(k) = (I-\bm{K}\bm{\xi}(k))\bm{\Gamma}(k-1)
    \label{eq:33}
\end{equation}

This regressor $\bm{\xi}$ consists of the body slip angle $\beta$. As a result, an adaptive observer was used to estimate the body slip angle ($\hat{\beta}$). It is referred to as an adaptive observer because the coefficients of the state equations presented below (equations \eqref{eq:34} and \eqref{eq:35}) are dependent on cornering stiffnesses and are updated at each time step by the values provided by the cornering stiffness estimator (equation \eqref{eq:31}).

\begin{equation}
    \bm{\hat{\dot x} = A(\hat{\theta}) \hat{x} + B(\hat{\theta})(y-\hat{y})}
    \label{eq:34}
\end{equation}

\begin{equation}
    \bm{\hat{y} = C(\hat{\theta})\hat{x} + D(\hat{\theta})u}
    \label{eq:35}
\end{equation}

Here, $\bm{\hat{x}}$ and $\bm{\hat{y}}$ are the estimated state vector and estimated output vector. $\bm{u}$ is the input vector. These three vectors are defined as follows.

\begin{equation}
    \bm{x} = \begin{bmatrix} \beta \\ \gamma \end{bmatrix} \hspace{5pt}
    \bm{y} = \begin{bmatrix} \gamma \\ a_y \end{bmatrix} \hspace{5pt}
    \bm{u} = \begin{bmatrix} \delta \\ N_z \end{bmatrix}
    \label{eq:36}
\end{equation}

The state equations of the observer are based on the Eqs. \eqref{eq:21}, \eqref{eq:22} and \eqref{eq:23}. $\bm{A}$ is the state transition matrix, $\bm{B}$ is the input matrix, $\bm{C}$ is the output matrix and $\bm{D}$ is the direct transition matrix which is defined as, 

\begin{equation}
    \bm{A}(\theta) = 
    \begin{bmatrix}
        -2\left(\mfrac{C_f+C_r}{Mv}\right) & -1-2\left(\mfrac{l_f C_f - l_r C_r}{Mv^2}\right) \\[1em]

        -2\left( \mfrac{l_f C_f-l_r C_r}{I} \right) & -2\left( \mfrac{l_f^2 C_f - l_r^2 C_r}{Iv} \right)
    \end{bmatrix}
\end{equation}

\begin{equation}
    \bm{B}(\theta) = 
    \begin{bmatrix}
    \mfrac{2C_f}{Mv} & 0 \\[1em]
    \mfrac{2 l_f C_f}{I} & \mfrac{1}{I}
    \end{bmatrix}
\end{equation}

\begin{equation}
    \bm{C}(\theta) = 
    \begin{bmatrix}
        0 & 1 \\
        v a_{11} & v(a_{12}+1)
    \end{bmatrix}
\end{equation}

\begin{equation}
     \bm{D}(\theta) = 
    \begin{bmatrix}
        0 & 0 \\
        v b_{11} & 0
    \end{bmatrix}
\end{equation}

where $a_{ij}$ and $b_{ij}$ are the elements of \bm{$A$} and \bm{$B$} matrices. 

As shown in Figure \ref{TSC}(a), the estimated cornering stiffnesses are used to calculate the yaw moment due to lateral forces using equation \eqref{eq:23} and compensate for it when the corrective yaw moment is generated. Furthermore, a PI controller has been attached to generate a control yaw signal to track the desired yaw rate.

The estimation error can be represented as $e =\beta - \hat{\beta}$ and $\dot e = (\bm{A}-\bm{K}\bm{C})e $. For robust performance, the gain matrix should be selected properly. All eigenvalues of $\bm{(A - KC)}$ must be negative to ensure the stability of the observer. Based on the pole assignment and robustness, the gain matrix $\bm{K}(\theta)$ was derived as,

\begin{equation}
    \bm{K}(\theta) = 
    \begin{bmatrix}
        \mfrac{\lambda_1 \lambda_2(l_f C_f - l_r C_r) I}{(2C_f C_r (l_f + l_r)^2)} -1 & \mfrac{1}{v} \\
        -(\lambda_1 + \lambda_2) & \mfrac{M ( l_f^2 C_f + l_r^2 C_r )}{I(l_fC_f + l_r C_r)}
    \end{bmatrix}
    \label{eq:41}
\end{equation}

where, $\lambda_1$ and $\lambda_2$ are the assigned poles of the observer.

\subsection{Differential Action}
During cornering, the wheels rotate at different speeds because each wheel has its own turning radius. The turning radius of the vehicle can be obtained using the steering angle ($\delta$) and yaw rate ($\gamma$) as follows,

\begin{equation}
    R = L/tan(\delta)
    \label{eq:42}
\end{equation}

The radii of rear left and rear right wheels ($R_l$ and $R_r$) can be derived as ${R_l} = R+t/2$ and ${R_r} = R-t/2$, where $t$ is the track width. The commanded voltage on the motors ($V_{cmd}$) was considered as a function of the torque demand of the driver ($T_{dem}$) that is sensed by the accelerator pedal position sensor. Then $V_{cmd}$ was modified based on the turning radii of each wheel to generate the individual commanded voltages for left and right wheel motors ($V_{cmd,l}$,$V_{cmd,r}$ respectively).

\begin{equation}
    {V_{cmd,l}} = \mfrac{R_l}{R} {V_{cmd}}
    \hspace{10pt}
    {V_{cmd,r}} =  \mfrac{R_r}{R} {V_{cmd}}
    \label{eq:44}
\end{equation}

It should be noted that the turning radii of the two wheels are not the same as the values given by $R_l$ and $R_r$ expressions when the vehicle is running. This is due to non-zero slip angles generated by each wheel due to lateral tire forces. Thus, the radii must be modified depending on the cornering stiffnesses. But in this context, it is assumed that the error of cornering radii is small enough since the yaw rate is controlled in such a way that slip angles hardly exceed the \ang{10} range.   

\subsection{Controller Integration using Fuzzy Inference System (FIS)}

\begin{table}[tbp]
  \centering
 \caption{Fuzzy rules for controller integration}
\begin{threeparttable}
  \setlength{\tabcolsep}{0.6em}
    \begin{tabular}{cccccccccccc}
    \toprule
    \multicolumn{2}{c}{\multirow{3}[5]{*}{$V_{corr,l}$}} & \multicolumn{10}{c}{Yaw Rate Error ($\gamma_{err}$)} \\
    \cmidrule(l){3-12}    \multicolumn{2}{c}{} & NL  & NS  & Z   & PS  & PL  & NL  & NS  & Z   & PS  & PL \\
    \cmidrule(lr){3-7} \cmidrule(l){8-12}     \multicolumn{2}{c}{} & \multicolumn{5}{c}{Left Motor} & \multicolumn{5}{c}{Right Motor} \\
    \cmidrule(r){1-2} \cmidrule(lr){3-7} \cmidrule(l){8-12}
    \multirow{5}[2]{*}{\rotatebox{90}{Slip} \newline \rotatebox{90}{Ratio}} & VS  & PL  & PL  & Z   & NL  & NL  & NL  & NL  & Z   & PL  & PL \\
        & S   & PL  & PS  & N   & NS  & NL  & NL  & NS  & N   & PS  & PL \\
        & M   & PS  & Z   & NS  & NL  & NL  & NL  & NL  & NS  & Z   & PS \\
        & L   & PS  & Z  & NL   & NS  & NL  & NL  & NS  & NL   & Z  & PS \\
        & VL   & Z  & Z   & NL  & NL  & NL  & NL  & NL  & NS  & Z   & Z \\
    \bottomrule
     \end{tabular}
    Normalised corrective voltage ($V_{corr}$) and yaw rate error ($\gamma_{err}$) fuzzy sets: Negative Large: NL, Negative Small: NS, Zero: 0, Positive Small: PS, Positive Large: PL. 
    Normalised slip ratio ($\lambda$) fuzzy sets: VS: Very Small, S: Small, M: Medium, L: Large, and VL: Very Large.
  \end{threeparttable}
    \label{tab:fuzzy_rules}
\end{table}%

\begin{figure}[tbp]
\begin{center}
\vspace{-10pt}
\includegraphics[width=0.5\textwidth]{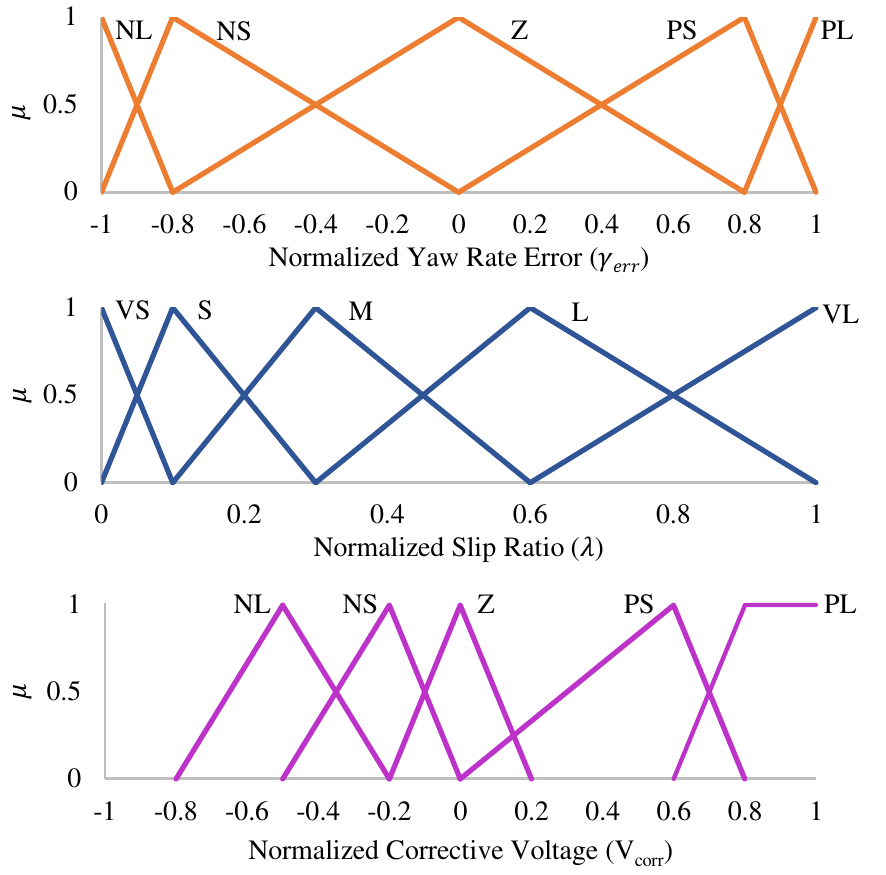}
\caption{Input and output membership functions of the fuzzy inference system (FIS): (a) Normalised yaw rate error ($\gamma_{err}$), (b) Normalised slip ratio ($\lambda$) and (c) Normalised corrective voltage ($V_{corr}$).}
\label{MFs}
\end{center}
\vspace{-25pt}
\end{figure}

The FIS generates modifications needed for the commanded voltages using correction factors so that the traction and yaw rate are controlled as expected by the individual controllers. The inputs to the FIS are slip ratio ($\lambda$) and yaw rate error ($\gamma_{err}$) which is defined as follows. 

\begin{equation}
    \gamma_{err} = \gamma - \gamma_{des}
    \label{eq:45}
\end{equation}

The voltage correction factors for left and right wheel motors are $V_{corr,l}$ and $V_{corr,r}$ respectively. All variables except for slip ratio were normalized and fuzzified into five fuzzy sets namely, NL - negative large, NS - negative small, Z - zero, PS - positive small, and PL - positive large. Since only positive and zero slip ratio values were considered (e.i non-braking situations) a separate set of five fuzzy sets were used: VS - Very Small, S - Small, M - Medium, L - Large, VL - Very Large. Triangular membership functions were used for all four variables as shown in Figure \ref{MFs}. Min and Max functions were employed in fuzzy rules for AND and OR operations. A Mamdani-type fuzzy inference system was employed with the Min implication and Max aggregation methods. The centroids of the inferred membership functions were used for defuzzification. Table \ref{tab:fuzzy_rules} shows the rule base for the right and left motors. The corrective yaw moment is generated by the difference between the correction factors of the two wheels. When a non-trivial yaw rate error is identified, the correction factor for one wheel increases while the other decreases. However, when a high slip ratio is detected, both correction factors are shifted more towards the negative values to make sure that the magnitude of the positive correction factor is as small as possible, but the torque difference is enough to correct the yaw moment at the same time.

\section{Model Implementation and Experiments}
    \label{Experiments}

\begin{table}[tbp]
\renewcommand{\arraystretch}{0.5}
\setlength{\tabcolsep}{0.6em}
  \centering
  \caption{\small Simulation model parameters}
  \subcaption{Dimensional parameters of the vehicle used in the simulation model.}
  \vspace{-5pt}
    \begin{tabular}{ccccccc}
    \toprule
    M (kg) & I\textsubscript{z} (kgm\textsuperscript{2}) & l\textsubscript{f} (m) & l\textsubscript{r} (m) & w (m) & J\textsubscript{w} (kgm\textsuperscript{2}) & d (m) \\
    \midrule
    260 & 60  & 0.83 & 0.7 & 1.2 & 0.23 & 0.46 \\
    \bottomrule
    \end{tabular}%
    \label{tab:vehicle parameters}%
    
\vspace{5pt}
\subcaption{Motor parameters used in the simulation model.}
\vspace{-5pt}
\renewcommand{\arraystretch}{0.5}
\begin{tabular}{ccccccc}
    \toprule
    \makecell{J\textsubscript{m} \\ (kgm\textsuperscript{2})}&  \makecell{K\textsubscript{t} \\ (NmA\textsuperscript{-1})} &  \makecell{K\textsubscript{f} \\ (Nmrad\textsuperscript{-1}s)} &  \makecell{K\textsubscript{E} \\(Vrad\textsuperscript{-1}s)} &  \makecell{R \\ ($\Omega$)} &  \makecell{L \\ (H)} &  \makecell{Pole \\ pairs} \\
    \midrule
    1.26e-2 & 0.5  & 0.01 & 0.04 & 7.0e-3 & 7.6e-5 & 10 \\
    \bottomrule
    \end{tabular}%
  \label{tab:motor parameters}%
  \vspace{-10pt}
\end{table}

\begin{figure}[tbp]
\begin{center}
\includegraphics[width=0.5\textwidth]{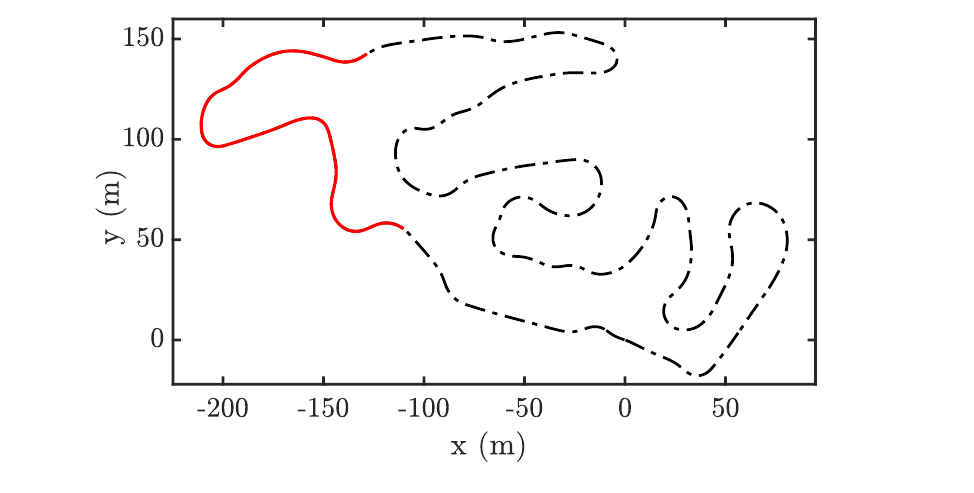}
\vspace{-20pt}
\caption{FSG Hockenheimring race track: Segment marked in red was used for the simulation.}
\vspace{-25pt}
\label{fig:Track}
\end{center}
\end{figure}

This section describes the simulation model that was used to implement the EDTSC and the experimental method followed to assess the performance of EDTSC.

\subsection{Simulation Model}
The EDTSC was integrated to a twin-track non-linear transient-state vehicle model with non-linear tire dynamics. Aerodynamic forces were added to the model to improve the accuracy of the simulation. Parameters of a formula-type electric racecar was used for the simulation. The parameters are given in table \ref{tab:vehicle parameters}. A permanent magnet synchronous motor model was used as the power-plant for the vehicle model and its parameters are given in table \ref{tab:motor parameters}. A simplified version of a pacejka tire model \cite{pacejka2006tyre} was used to simulate tire dynamics. 

The 6-DOF model was implemented MATLAB and solved numerically using the MATLAB ode23 ordinary differential equation solver. Furthermore, a constant time step of 5\unit{ms} was used. All simulations were performed on a laptop with an i7-10750H processor (2.60 GHz) and 8GB of RAM running on Windows 10.


\subsection{Experiments}
A predefined track using an adaptive driver model and double lane change by following a reference path was used to validate the proposed EDTSC in a simulated environment. In each scenario, the initial phase of testing was done to confirm that the cornering stiffness estimator, reaction force observer and slip ratio estimator are closely estimating the parameters on which the EDTSC rely on. For track simulation, the Formula Student Germany (FSG) Hockenheimring endurance track was used \cite{FSG}. For simplicity, only the results of a segment (represented by red in Figure \ref{fig:Track}) are discussed in the next section \ref{Results}. The driver model was fed with a series of curvature-distance data, and its task was to keep the path as close to the prescribed curvature as possible. The initial vehicle velocity was set to zero. 

The ISO 3888-2:2011 standard \cite{ISO3888-2}  was used to define the double lane change path. The path was tracked by the driver model in the same way as in the previous case, with the exception that the driver let off the throttle at the start of the path after reaching the test velocity. The controller was put through its paces at two different longitudinal speeds: 40 \unit{km.h^{-1}} and 100 \unit{km.h^{-1}}.

\section{Results}
    \label{Results}
This section presents the numerical results obtained when the EDTSC architecture presented in section \ref{EDTSC System Development} is applied to the simulation model in section \ref{Experiments}.

\subsection{Race Track Simulation}

\begin{figure}[!t]
\begin{center}
    \includegraphics[width=0.48\textwidth]{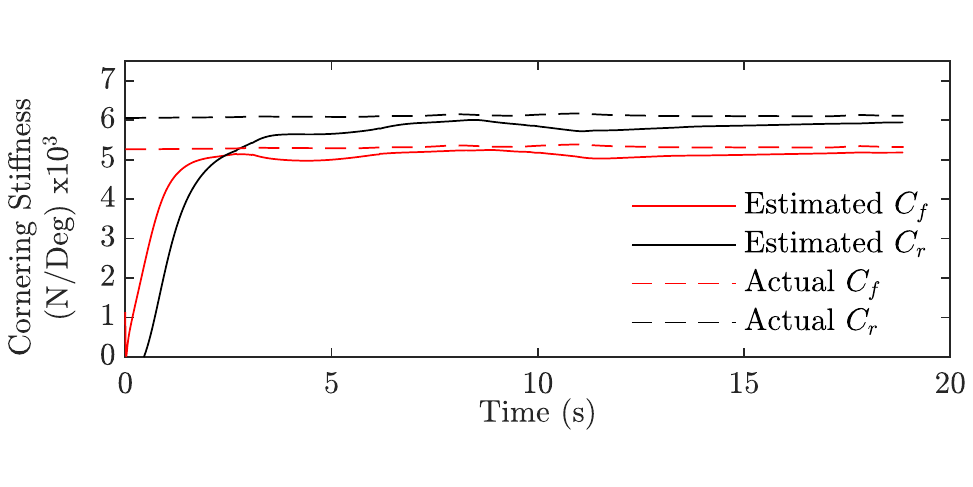}
    \vspace{-15pt}
    \caption{Variation of estimated front and rear cornering stiffnesses ($C_f$ and $C_r$) with time for the race track simulation.}
    \label{fig:Cornering Stiffness}
    \vspace{-20pt}
    \end{center}
\end{figure}

\begin{figure*}[!t]
\begin{center}
    \includegraphics[width=\textwidth]{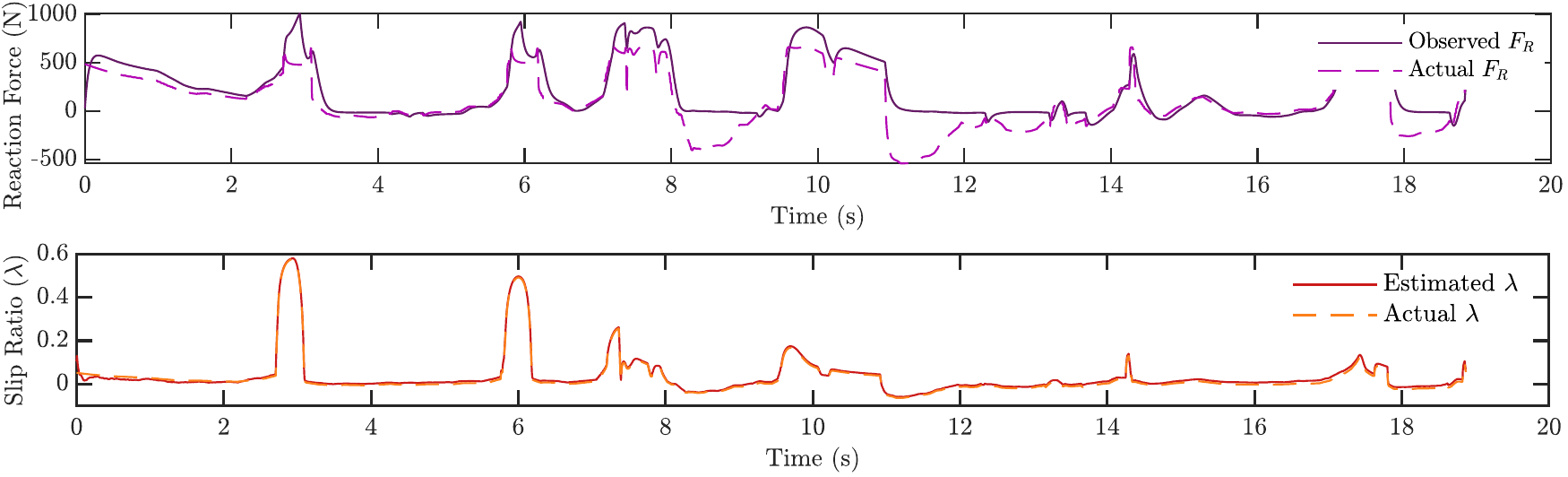}
    \vspace{-15pt}
    \caption{Reaction force (top) and Slip ratio (bottom) estimates against time.}
    \label{fig:Actual vs Estimated}
    \vspace{-20pt}
    \end{center}
\end{figure*}

\begin{figure*}[htbp]
\begin{center}
\includegraphics[width=\textwidth]{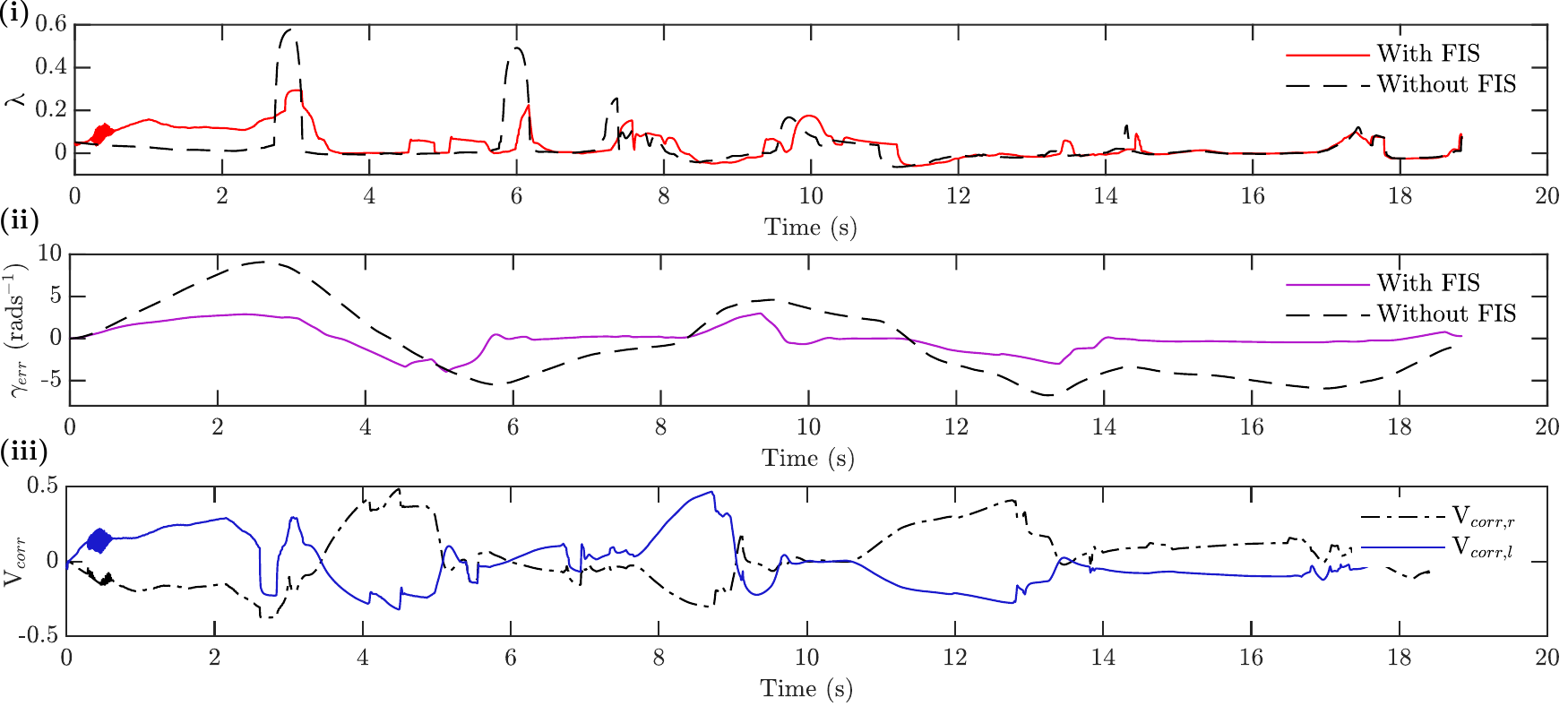}
\vspace{-15pt}
\caption{Race track simulation results against time: (i) - Variation of slip ratio ($\lambda$), (ii) - Variation of yaw rate error ($\gamma_{err}$), with and without fuzzy inference system (FIS), (iii) - Variation of normalized corrective voltage factors ($V_{corr}$) generated by the FIS.}
\vspace{-20pt}
\label{fig:With FIS vs Without FIS}
\end{center}
\end{figure*}

The accuracy with which the cornering stiffness estimator estimates the actual cornering stiffnesses are shown in Figure \ref{fig:Cornering Stiffness}. It is evident that the estimator is capable of reaching the actual cornering stiffness values faster. 

The top graph of Figure \ref{fig:Actual vs Estimated} depicts actual reaction force ($F_R$) and observed reaction force variations against time. It can be seen that for positive reaction force values, the reaction force observer tracks the actual force. However, a significant deviation can be observed at negative values because the reaction force observer does not account for braking situations. This is not a problem because the proposed FIS copes with non-braking  situations only. In accelerating situations, the observer can be seen tracking the actual value accurately with a correlation coefficient $> 0.85$. The slip ratio ($\lambda$) estimation is accurate, and the error is negligible (with a correlation coefficient $>0.99$) as shown in the bottom graph of Figure \ref{fig:Actual vs Estimated}. This suggests that slip ratio can be safely used in the FIS.

\begin{table}[tbp]
  \centering
  \caption{\small Simulation Results}
  \begin{threeparttable}
\renewcommand{\arraystretch}{0.8}
  \setlength{\tabcolsep}{0.5em}
    \begin{tabular}{cccc}
    \toprule
    \begin{tabular}[x]{@{}c@{}}Simulation\\type\end{tabular} & Parameter & \begin{tabular}[x]{@{}c@{}}RMS value\\reduction (\%)\end{tabular} & \begin{tabular}[x]{@{}c@{}}Peak value\\reduction (\%)\end{tabular} \\
    \midrule
    \multirow{2}[1]{*}{\begin{tabular}[x]{@{}c@{}}Race\\track\end{tabular}} & $\lambda$ & 96.14 & 42.31 \\
        & $\gamma_{err}$ & 88.17 & 71.93 \\
    \midrule
    \multirow{2}[1]{*}{\begin{tabular}[x]{@{}c@{}}Double lane\\change\end{tabular}} & $\gamma_{err}$ (40 \unit{km.h^{-1}}) & 86.96 & 97.93 \\
        & $\gamma_{err}$ (100 \unit{km.h^{-1}}) & 92.34 & 120.24 \\
    \bottomrule
    \end{tabular}
    $\lambda$: Slip ratio, $\gamma_{err}$: Yaw rate error
    \end{threeparttable}
  \label{tab:simulation results}
  \vspace{-10pt}
\end{table}

Figure \ref{fig:With FIS vs Without FIS} depicts the resulting slip ratio ($\lambda$), yaw rate error ($\gamma_{err}$), and FIS output ($V_{corr}$) variations. Figure \ref{fig:With FIS vs Without FIS}-(i) shows the slip ratio improvement with the FIS. The FIS has flattened out the peaks caused by the sudden increase in wheel velocities. Figure \ref{fig:With FIS vs Without FIS}-(ii) depicts the compensation for yaw rate error, and it shows how a lower error could be achieved throughout the run time compared to the case without the FIS. The percentage reductions in slip ratio and yaw rate error for track simulation are shown in Table \ref{tab:simulation results}. Another significant finding is that the total time required by the vehicle to complete a track segment with FIS ($18.845$ \unit{s}) is less than that required without it ($18.85$ \unit{s}). This indicates that the implemented EDTSC system does not degrade the vehicle's overall performance in exchange for improved stability, but rather the opposite. This $0.3\%$ time improvement is significant in terms of racecar lap times where racecars have to complete multiple laps.

As shown in Figure \ref{fig:With FIS vs Without FIS}-(iii), the values of the corrective voltage factors are biased towards positive values due to the nature of the membership functions. This ensures that the motor control saturation caused by the generation of negative voltage commands is limited, thereby increasing the range of the corrective yaw moment.

\subsection{Double Lane Change Simulation}

 \begin{figure*}[tbp]
 \begin{center}
\includegraphics[width=\textwidth]{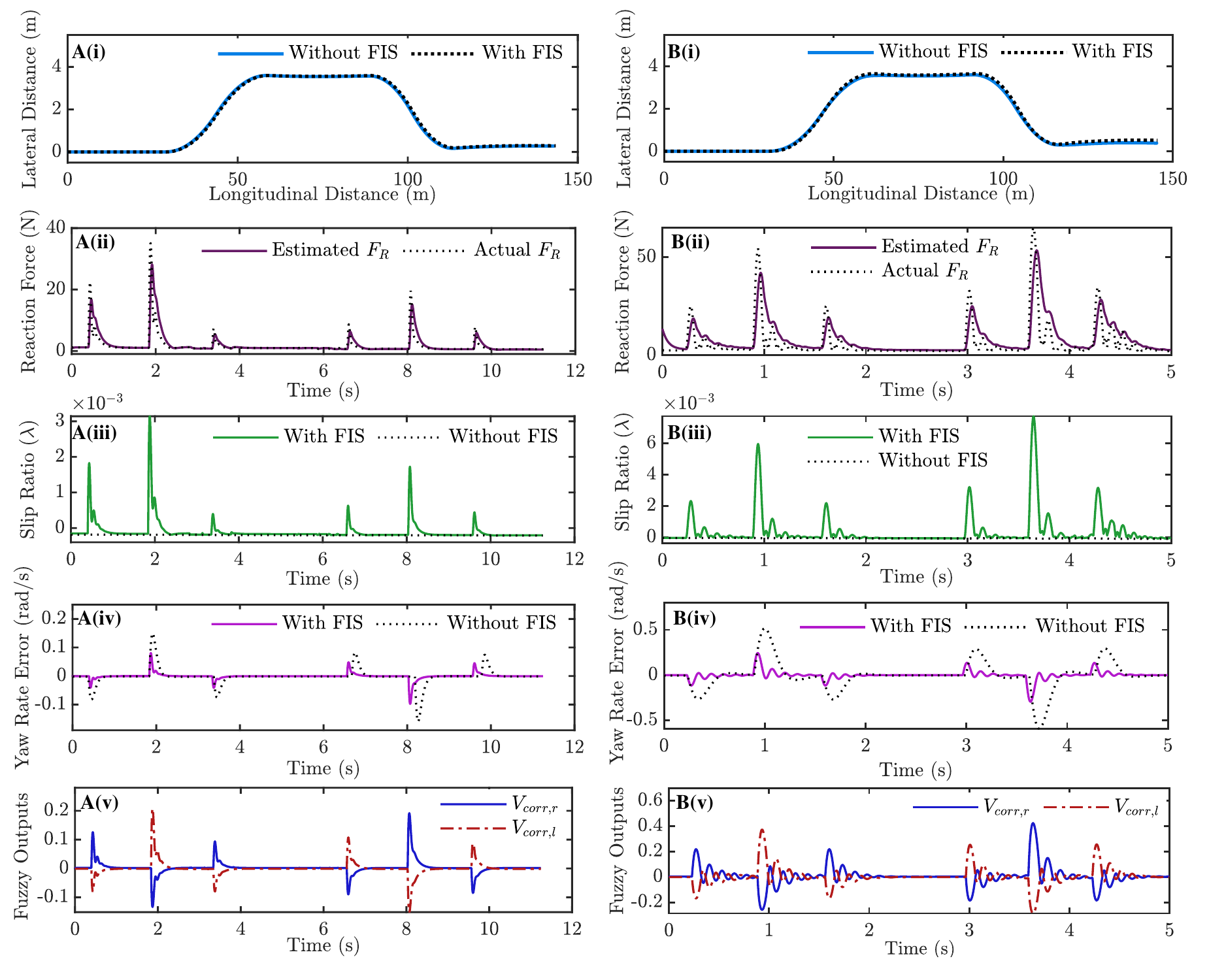}
 \vspace{-25pt}
 \caption{Double lane change manoeuvre simulation results with with and witout Fuzzy Inference System (FIS): A(i) - A(iv) for 40 \unit{km.h^{-1}} and B(i) - B(iv) for 100 \unit{km.h^{-1}}.}
 \vspace{-25pt}
 \label{DLC}
 \end{center}
 \end{figure*}

The paths traced by the vehicle with and without the FIS are shown in \ref{DLC} A(i) and B(i). It is clear that the addition of the FIS has not compromised path tracking, and it will be demonstrated in the following that stability has improved. Although the driver's throttle input is zero during the double lane change manoeuvre, the FIS is permitted to generate non-zero motor voltage commands to ensure that the yaw rate is as close to the desired value as possible. This explains why the curves in \ref{DLC} A(ii) and B(ii) show non-zero reaction forces. The correlation between the observed reaction force and actual reaction force is $>0.8$ for 40 \unit{km.h^{-1}} and $> 0.75$ for 100 \unit{km.h^{-1}}. As a result of the FIS's behaviour, the slip ratio becomes positive at each location where the vehicle attempts to negotiate a curve, as illustrated in \ref{DLC} A(iii) and B(iii). However, it is clear that when the vehicle is run without the FIS, the slip ratio is negative and less than $2\times10^{-4}$, whereas it increases at the beginning of curved parts of the path when the FIS is used. The root mean square slip ratios for 40 \unit{km.h^{-1}} test is $< 0.001$ and for 100 \unit{km.h^{-1}} test it is $<0.002$. Because the peak slip ratio is within the linear region of the tyre longitudinal force ($\lambda <0.04$ for 40 \unit{km.h^{-1}}, $\lambda <0.08$ for 100 \unit{km.h^{-1}}), this sacrifice in low slip ratio is acceptable.

From Figure \ref{DLC} A(iv) and B(iv) it can be concluded that the FIS performs well in reducing the yaw rate error relative to the desired yaw rate. The reductions in root mean square and peak values of yaw rate error are given in Table \ref{tab:simulation results}.

It was also observed that increase in FIS gain could lead to oscillatory behaviour of yaw rate thus the gain value must be carefully selected otherwise it would result in instability. However, the selection of FIS gain considering the stability of the controller is out of the scope of this research.
The oscillatory behaviour origins from the corrective voltage factors that are generated by the FIS and is evident from \ref{DLC} B(v). As shown in Figure \ref{DLC} A(v) and B(v), the corrective voltage factors suddenly goes up at high yaw rate error points (during cornering) to generate the corrective yaw moment. As in the track simulation case, both factors are biased towards positive to avoid quick zero saturation of motor torque.

\section{Conclusion}
    \label{Conclusion}
\vspace{5pt}
This article proposes a novel controller for electronic differentials with an FIS to integrate the direct torque and direct yaw moment controllers. The controller was integrated with a 6-DOF vehicle model and simulated for two test cases. The slip ratio and cornering stiffness values were estimated successfully to generate EDTSC reference parameters.To construct the cornering stiffness estimator's regressor in the RLS algorithm, a slip angle observer based on the bicycle model was used. The slip ratio was estimated closely to the actual value using the reaction torque observer that is based on a disturbance observer. 

To reduce dynamical instabilities, the FIS could handle the trade-off between slip control and yaw rate control by sending appropriate control signals to the traction motors. Accurate enough slip reaction force and slip ratio estimations could be achieved in various drive conditions. Furthermore, the FIS could secure a substantial reduction in yaw rate error relative to the desired yaw rate and mitigate the peak slip ratios resulted from sudden change in driving maneouver. The compromise in slip ratio during double lane change test was negligible. Future work could focus on implementing the EDTSC in a test vehicle to evaluate the performance. In addition, four-wheel electric drive vehicles could be considered when developing the control architecture to improve traction and stability.


\vspace{5pt}

\section*{Acknowledgment}
    The authors thank the Falcon E Racing, the electric formula student team of the University of Moratuwa for the insightful discussions. Also, the authors acknowledge the technical and financial sponsors of the formula student project. 
\vspace{-3pt}

\newcommand{\BIBdecl}{\setlength{\itemsep}{-0.01 em}}
\bibliographystyle{IEEEtran}
\bibliography{References}

\end{document}